\documentclass{elsart}

\usepackage{graphics}
\usepackage{graphicx,float}
\usepackage{epsfig}
\usepackage{amsmath,amsfonts,amsgen}
\usepackage{dcolumn}
\usepackage[all]{xy}
\usepackage{makeidx,color}
\usepackage{color}

\newcolumntype{d}[1]{D{.}{\cdot}{#1}}
\newcolumntype{.}{D{.}{.}{-1}}

\newcommand\tabcaption{\def\@captype{table}\caption}
\newcommand{\be}{\begin{eqnarray*}}
\newcommand{\ee}{\end{eqnarray*}}
\newcommand{\ibe}{\begin{eqnarray}}
\newcommand{\iee}{\end{eqnarray}}

\def\3hf{\frac{3}{2}}

\def\np1{n+1}

\def\ip3hf{i+\frac{3}{2}}
\def\jp3hf{j+\frac{3}{2}}

\begin{document}
\begin{frontmatter}

\title{The scaling-law flows: An attempt at scaling-law vector calculus}
\author{Xiao-Jun Yang$^{1,2,3,\ast}$}
\corauth[cor1]{Corresponding author: Tel.: 086-67867615; E-Mail: dyangxiaojun@163.com (X. J. Yang)}

\address{$^{1}$ State Key Laboratory for Geomechanics and Deep Underground Engineering,
China University of Mining and Technology, Xuzhou 221116, China}
\address{$^{2}$ School of Mathematics, China University of Mining and Technology, Xuzhou 221116, China}
\address{$^{3}$ School of Mechanics and Civil Engineering, China University of Mining and Technology, Xuzhou 221116, China}

\begin{abstract}
In this paper, the scaling-law vector calculus, which is related to the
connection between the vector calculus and the scaling law in fractal
geometry, is addressed based on the Leibniz derivative and Stieltjes
integral for the first time. The Gauss-Ostrogradsky-like theorem,
Stokes-like theorem, Green-like theorem, and Green-like identities are
considered in the sense of the scaling-law vector calculus. The Navier-Stokes-like
equations are obtained in detail. The obtained result is as a potentially
mathematical tool proposed to develop an important way of approaching this
challenge for the scaling-law flows.
\end{abstract}


\end{frontmatter}

\section{Introduction}
The classical calculus is called the Newton-Leibniz calculus, which contains
the differential calculus and the integral calculus. The differential calculus
was proposed by Newton in 1665 \cite{1,2} and by Leibniz in 1684 \cite{3}. The integral
calculus was coined by Newton in 1665 \cite{1,2} and by Leibniz in 1686 \cite{4}. Based
on the Newton-Leibniz calculus, the vector calculus, denoted by Hamilton in
1844 \cite{5}, by Tait in 1890 \cite{6}, by Heaviside in 1893 \cite{7}, and by Gibbs in
1901\cite{8}, was applied in the fields of mechanics, hydrodynamics, and
electricity \cite{9}.

The calculus with respect to monotone functions is one of the classes of the
general calculus operators \cite{10,11}. This theory consists of the differential
calculus with respect to monotone function, which is called the Leibniz
derivative due to Leibniz \cite{12,13}, and the integral calculus with respect to
monotone function, which is called the Stieltjes integral due to Stieltjes
\cite{14}. The integral calculus with respect to monotone function was developed
by Widder \cite{15}, by Horst \cite{16} and by Stoll \cite{17}, respectively. The vector
calculus with respect to monotone function was proposed in \cite{18}.

The scaling laws are the connections between the fractal geometry and
measure in various complex phenomena \cite{19,20}. The experimental evidence for
the flow in the extended self-similarity scaling laws was considered in
\cite{21}. The scaling laws for the turbulent flow in the pipes were presented in
\cite{22,23}. The scaling laws for the wall-bounded shear flows were developed in
\cite{24}. The self-similarity scaling laws in turbulent flows was discussed in
\cite{25}.

The scaling-law calculus, which is considered to develop the connection
between the fractal geometry and calculus with respect to monotone
functions, was proposed to model the anomalous rheology (see \cite{26}). The
Gauss \cite{27}, Ostrogradsky \cite{28}, Stokes \cite{29} and Green \cite{30} tasks have not
been extended in the sense of the scaling-law calculus. Due to the present
investigation for the scaling-law differential calculus and the scaling-law
integral calculus, the scaling-law vector calculus has not been developed
based on the vector calculus with respect to monotone function. Motivated by
the present idea, the aim of the present paper is to propose the definitions
for the scaling-law vector calculus, to present its fundamental theorems,
and to suggest the potential and important applications in scaling-law
flows. The structure of the paper is designed as follows. In Section 2,
the general calculus operators are given. In Section 3, the theory and
properties of the scaling-law vector calculus are presented. In Section 4,
the Navier-Stokes-type equations for the scaling-law flows are discussed.
Finally, the conclusions are given
in Section 5.

\section{Preliminaries}

In this section, we introduce the definitions and theorems of the general
calculus operators containing the calculus with respect to monotone function
and scaling-law calculus.

\subsection{The calculus with respect to monotone function }

Let $\varphi _\vartheta \left( t \right)=\left( {\varphi \circ \vartheta }
\right)\left( t \right)=\varphi \left( {\vartheta \left( t \right)}
\right)$, where $\vartheta \left( t \right)$ is the monotone function, e.g.,
$\vartheta ^{\left( 1 \right)}\left( t \right)=d\vartheta \left( t
\right)/dt>0$.

Let $\Lambda \left( \varphi \right)$ be the set of the continuous
derivatives of the functions $\varphi \left( \vartheta \right)$ with respect
to the variable $\vartheta $ in the domain $\Im $.

Let $\Xi \left( \vartheta \right)$ be the set of the continuous derivatives
of the functions $\vartheta \left( t \right)$ with respect to the variable
$t$ in the domain $\aleph $.

Let us consider the set of the continuous derivatives of the composite
functions, defined as follows:
\[
\Re \left( {\varphi _\vartheta } \right)=\left\{ {\varphi _\vartheta \left(
t \right):\varphi _\vartheta \left( t \right)\in \Lambda \left( \varphi
\right),\vartheta \in \Xi \left( \vartheta \right)} \right\}.
\]

\subsection{The Leibniz derivative}

Let $\varphi _\vartheta \in \Re \left( {\varphi _\vartheta } \right)$. The
Leibniz derivative of the function $\varphi _\vartheta \left( t \right)$ is
defined as \cite{11,18,26}
\begin{equation}
\label{eq1}
D_{t,\vartheta }^{\left( \mbox{1} \right)} \varphi _\vartheta \left( t
\right)=\frac{1}{\vartheta ^{\left( 1 \right)}\left( t
\right)}\frac{d\varphi _\vartheta \left( t \right)}{dt}.
\end{equation}
The geometric interpretation of the Leibniz derivative is the rate of change
of the functional $\varphi _\vartheta \left( t \right)$ with the function
$\vartheta \left( t \right)$ in the independent variable $t$ \cite{11,18,26}.

Let $\varphi _\vartheta \in \Re \left( {\varphi _\vartheta } \right)$. The
total Leibniz-type differential with respect to monotone function $\vartheta
\left( t \right)$ of the function $\varphi _\vartheta \left( t \right)$,
denoted as $d\varphi _\vartheta \left( t \right)=d\varphi \left( {\vartheta
\left( t \right)} \right)$, is defined as
\begin{equation}
\label{eq2}
d\varphi _\vartheta \left( t \right)=\left( {\vartheta ^{\left( 1
\right)}\left( t \right)D_{t,\vartheta }^{\left( {1} \right)} \varphi
_\vartheta \left( t \right)} \right)dt.
\end{equation}

\subsection{The Stieltjes integral }

Let $\Phi _\vartheta \in \Re \left( {\Phi _\vartheta } \right)$. The
Stieltjes integral of the function $\Phi _\vartheta \left( t \right)$ in the
interval $\left[ {a,b} \right]$ is defined as \cite{11,18,26}
\begin{equation}
\label{eq3}
{ }_aI_{b,\vartheta }^{\left( {1} \right)} \Phi _\vartheta \left( t
\right)=\int\limits_a^b {\Phi _\vartheta \left( t \right)\vartheta ^{\left(
1 \right)}\left( t \right)dt} .
\end{equation}
Similarly, the geometric interpretation of the Stieltjes integral is the
area enclosed by the integrand function $\Phi _\vartheta \left( t \right)$
and the function $\vartheta \left( t \right)$ in the independent variable
$t\in \left[ {a,b} \right]$ \cite{11,18,26}.

Their properties are given as follows:

(${\rm O}1)$ The chain rule for the Leibniz derivative is given as follows
\cite{18}:
\begin{equation}
\label{eq4}
D_{t,\vartheta }^{\left( {1} \right)} \Theta \left\{ {\varphi
_\vartheta \left( t \right)} \right\}=\Theta ^{\left( 1 \right)}\left(
\varphi \right)\cdot D_{t,\vartheta }^{\left( {1} \right)} \varphi
_\vartheta \left( t \right),
\end{equation}
where $\Theta ^{\left( 1 \right)}\left( \varphi \right)=d\Theta \left(
\varphi \right)/d\varphi $.

(${\rm O}2)$ The change-of-variable theorem for the Stieltjes integral reads
as follows \cite{18}:
\begin{equation}
\label{eq5}
{ }_aI_{t,\vartheta }^{\left( {1} \right)} \left( {\Theta ^{\left( 1
\right)}\left( \varphi \right)\cdot D_{t,\vartheta }^{\left( {1}
\right)} \varphi _\vartheta \left( t \right)} \right)=\Theta \left\{
{\varphi _\vartheta \left( t \right)} \right\}-\Theta \left\{ {\varphi
_\vartheta \left( a \right)} \right\}.
\end{equation}

\subsection{The Leibniz-type partial derivatives}

Let $\Theta =\Theta \left( {x,y,z} \right)=\Theta \left( {\alpha \left( x
\right),\beta \left( y \right),\gamma \left( z \right)} \right)$,
where $\alpha ^{\left( 1 \right)}\left( x \right)>0$, $\beta ^{\left( 1
\right)}\left( y \right)>0$ and $\gamma ^{\left( 1 \right)}\left( z
\right)>0$.

The Leibniz-type partial derivatives of the scalar field $\Theta $ are
defined as \cite{18}
\begin{equation}
\partial _{x,\alpha }^{\left( 1 \right)} \Theta =\frac{1}{\alpha ^{\left( 1
\right)}\left( x \right)}\frac{\partial \Theta }{\partial x},
\end{equation}

\begin{equation}
\partial _{y,\beta }^{\left( 1 \right)} \Theta =\frac{1}{\beta ^{\left( 1
\right)}\left( y \right)}\frac{\partial \Theta }{\partial y}
\end{equation}
and
\begin{equation}
\partial
_{z,\gamma }^{\left( 1 \right)} \Theta =\frac{1}{\gamma ^{\left( 1
\right)}\left( z \right)}\frac{\partial \Theta }{\partial z},
\end{equation}
respectively.

The total Leibniz-type differential of the scalar field $\Theta $ is defined
as \cite{18}:
\begin{equation}
\label{eq6}
d\Theta =\left( {\alpha ^{\left( 1 \right)}\left( x \right)\partial
_{x,\alpha }^{\left( 1 \right)} \Theta } \right)dx+\left( {\beta ^{\left( 1
\right)}\left( y \right)\partial _{y,\beta }^{\left( 1 \right)} \Theta }
\right)dy+\left( {\gamma ^{\left( 1 \right)}\left( z \right)\partial
_{z,\gamma }^{\left( 1 \right)} \Theta } \right)dz.
\end{equation}
which leads to
\begin{equation}
\label{eq7}
\frac{d\Theta }{dt}=\left( {\alpha ^{\left( 1 \right)}\left( x
\right)\partial _{x,\alpha }^{\left( 1 \right)} \Theta }
\right)\frac{dx}{dt}+\left( {\beta ^{\left( 1 \right)}\left( y
\right)\partial _{y,\beta }^{\left( 1 \right)} \Theta }
\right)\frac{dy}{dt}+\left( {\gamma ^{\left( 1 \right)}\left( z
\right)\partial _{z,\gamma }^{\left( 1 \right)} \Theta }
\right)\frac{dz}{dt}.
\end{equation}

\subsection{The scaling-law calculus}

Let us consider the set of the continuous derivatives of the composite
functions, defined as follows:
\[
\Re \left( {\varphi _\eta } \right)=\left\{ {\varphi _\vartheta \left( t
\right):\varphi _\vartheta \left( t \right)\in \Lambda \left( \varphi
\right),\vartheta \in \Xi \left( \vartheta \right)} \right\},
\]
where the fractal scaling law is defined as \cite{19,20}
\begin{equation}
\label{eq8}
\vartheta \left( t \right)=\lambda t^\eta
\end{equation}
with the normalization constant $\lambda \ge 0$, the radius $t\ge 0$, and
the scaling exponent $\eta \ge 0$.

Here, we take $-\infty <t<\infty $, $-\infty <\lambda <\infty $ and $-\infty
<\eta <\infty $.

\subsection{The scaling-law derivative}

Let $\varphi _\eta \in \Re \left( {\varphi _\eta } \right)$, e.g.,$\varphi
_\mu \left( t \right)=\left( {\varphi \circ \left( {\lambda t^\eta }
\right)} \right)\left( t \right)=\varphi \left( {\lambda t^\eta } \right)$.

The scaling-law derivative of the function $\varphi _\eta \left( t \right)$
is defined as \cite{19,26}
\begin{equation}
\label{eq9}
{ }^{SL}D_t^{\left( {1} \right)} \varphi _\eta \left( t
\right)=\frac{d\varphi _\eta \left( t \right)}{d\left( {\lambda t^\eta }
\right)}=\frac{1}{\lambda \eta t^{\eta -1}}\frac{d\varphi _\eta \left( t
\right)}{dt}.
\end{equation}
The geometric interpretation of the scaling-law derivative is the rate of
change of the functional $\varphi _\eta \left( t \right)$ with the function
$\vartheta =\lambda t^\eta $ in the independent variable $t$ \cite{19,26}.

Let $\varphi _\eta \in \Re \left( {\varphi _\eta } \right)$. The total
scaling-law differential of the function $\varphi _\eta \left( t \right)$,
denoted as $d\varphi _\eta \left( t \right)$, is defined as
\begin{equation}
\label{eq10}
d\varphi _\eta \left( t \right)={ }^{SL}D_t^{\left( {1} \right)}
\varphi _\eta \left( t \right)d\left( {\lambda t^\eta } \right)=\left(
{\lambda \eta t^{\eta -1}{ }^{SL}D_t^{\left( {1} \right)} \varphi _\eta
\left( t \right)} \right)dt.
\end{equation}

\subsection{The scaling-law integral }

Let $\Phi _\eta \in \Re \left( {\Phi _\eta } \right)$. The scaling-law
integral of the function $\Phi _\eta \left( t \right)$ in the interval
$\left[ {a,b} \right]$ is defined as \cite{19,26}
\begin{equation}
\label{eq11}
{ }_a^{SL} I_b^{\left( {1} \right)} \Phi _\eta \left( t
\right)=\int\limits_a^b {\Phi _\eta \left( t \right)d\left( {\lambda t^\eta
} \right)} =\int\limits_a^b {\Phi _\eta \left( t \right)\lambda \eta t^{\eta
-1}dt} .
\end{equation}
Similarly, the geometric interpretation of the scaling-law integral is the
area enclosed by the integrand function $\Phi _\eta \left( t \right)$ and
the function $\vartheta \left( t \right)=\lambda t^\eta $ in the independent
variable $t\in \left[ {a,b} \right]$ \cite{26}.

Their properties are presented as follows:

($P1)$ The chain rule for the scaling-law derivative is given as follows
\cite{19,26}:
\begin{equation}
\label{eq12}
{ }^{SL}D_t^{\left( {1} \right)} \Theta \left\{ {\varphi _\eta \left( t
\right)} \right\}=\Theta ^{\left( 1 \right)}\left( \varphi \right)\cdot {
}^{SL}D_t^{\left( {1} \right)} \varphi _\eta \left( t \right),
\end{equation}
where $\Theta ^{\left( 1 \right)}\left( \varphi \right)=d\Theta \left(
\varphi \right)/d\varphi $.

($P2)$ The change-of-variable theorem for the scaling-law integral can be
given as follows \cite{19,26}:
\begin{equation}
\label{eq13}
{ }_a^{SL} I_t^{\left({1} \right)} \left( {\Theta ^{\left( 1
\right)}\left( \varphi \right)\cdot { }^{SL}D_t^{\left( {1} \right)}
\varphi _\eta \left( t \right)} \right)=\Theta \left\{ {\varphi _\eta \left(
t \right)} \right\}-\Theta \left\{ {\varphi _\eta \left( a \right)}
\right\}.
\end{equation}

\subsection{The scaling-law gradient}

In order to discuss the scaling-law gradient, we consider the Cartesian-type
coordinate system $\left( {\lambda _1 x^{D_1 },\lambda _2 y^{D_2 },\lambda
_3 z^{D_3 }} \right)$, which leads to the Cartesian coordinate system
$\left( {x,y,z} \right)$, where the scaling exponents $D_1 =D_2 =D_3 =1$ and
$\lambda _1 =\lambda _2 =\lambda _3 =1$.

In the Cartesian-type coordinate system $\left( {\lambda _1 x^{D_1 },\lambda
_2 y^{D_2 },\lambda _3 z^{D_3 }} \right)$, the scaling-law gradient is
defined as
\begin{equation}
\label{eq14}
\begin{array}{l}
\nabla ^{\left( {{\begin{array}{*{20}c}
 {D_1 ,D_2 ,D_3 } \hfill \\
 {\lambda _1 ,\lambda _2 ,\lambda _3 } \hfill \\
\end{array} }} \right)}\\
=
i\left( {\lambda _1 D_1 x^{D_1 -1}} \right)\partial
_x^{\left( 1 \right)} +j\left( {\lambda _2 D_2 y^{D_2 -1}} \right)\partial
_y^{\left( 1 \right)} +k\left( {\lambda _3 D_3 z^{D_3 -1}} \right)\partial
_z^{\left( 1 \right)} ,
\end{array}
\end{equation}
where $i$, $j$ and $k$ are the unit vector in the Cartesian coordinate
system.

Let us consider the scaling-law scalar field, defined by:
\begin{equation}
\label{eq15}
X=X\left( {\lambda _1 x^{D_1 },\lambda _2 y^{D_2 },\lambda _3 z^{D_3 }}
\right).
\end{equation}
The scaling-law gradient of the scaling-law scalar field $X$ is given as
\begin{equation}
\label{eq16}
\begin{array}{l}
\nabla ^{\left( {{\begin{array}{*{20}c}
 {D_1 ,D_2 ,D_3 } \hfill \\
 {\lambda _1 ,\lambda _2 ,\lambda _3 } \hfill \\
\end{array} }} \right)}X\\
=
i\left( {\lambda _1 D_1 x^{D_1 -1}} \right)\partial
_x^{\left( 1 \right)} X+j\left( {\lambda _2 D_2 y^{D_2 -1}} \right)\partial
_y^{\left( 1 \right)} X+k\left( {\lambda _3 D_3 z^{D_3 -1}} \right)\partial
_z^{\left( 1 \right)} X.
\end{array}
\end{equation}
From (\ref{eq14}) and (\ref{eq15}) we have that
\begin{equation}
\label{eq17}
\begin{array}{l}
 dX\\
 =
 \left( {\lambda _1 D_1 x^{D_1 -1}} \right)\partial _x^{\left( 1 \right)}
Xdx+\left( {\lambda _2 D_2 y^{D_2 -1}} \right)\partial _y^{\left( 1 \right)}
Xdy+\left( {\lambda _3 D_3 z^{D_3 -1}} \right)\partial _z^{\left( 1 \right)}
Xdz \\
{ }=\nabla ^{\left( {{\begin{array}{*{20}c}
 {D_1 ,D_2 ,D_3 } \hfill \\
 {\lambda _1 ,\lambda _2 ,\lambda _3 } \hfill \\
\end{array} }} \right)}X\cdot {\rm {\bf n}}dl=\nabla ^{\left(
{{\begin{array}{*{20}c}
 {D_1 ,D_2 ,D_3 } \hfill \\
 {\lambda _1 ,\lambda _2 ,\lambda _3 } \hfill \\
\end{array} }} \right)}Xd{\rm {\bf l}}, \\
 \end{array}
\end{equation}
where ${\rm {\bf n}}$ is the unit normal to the surface, $dl$ is the
distance measured along the normal ${\rm {\bf n}}$, and $d{\rm {\bf l}}={\rm
{\bf n}}dl=idx+jdy+kdz$.

The scaling-law direction derivative of the scaling-law scalar field $X$
along the normal ${\rm {\bf n}}$ is defined as
\begin{equation}
\label{eq18}
\frac{dX}{dl}=\nabla ^{\left( {{\begin{array}{*{20}c}
 {D_1 ,D_2 ,D_3 } \hfill \\
 {\lambda _1 ,\lambda _2 ,\lambda _3 } \hfill \\
\end{array} }} \right)}X\cdot {\rm {\bf n}}=\partial _n^{\left(
{{\begin{array}{*{20}c}
 {D_1 ,D_2 ,D_3 } \hfill \\
 {\lambda _1 ,\lambda _2 ,\lambda _3 } \hfill \\
\end{array} }} \right)} X.
\end{equation}
The scaling-law Laplace-like operator, denoted as
\begin{equation}
\label{eq19}
\nabla ^{\left( {{\begin{array}{*{20}c}
 {2D_1 ,2D_2 ,2D_3 } \hfill \\
 {\lambda _1 ,\lambda _2 ,\lambda _3 } \hfill \\
\end{array} }} \right)}=\nabla ^{\left( {{\begin{array}{*{20}c}
 {D_1 ,D_2 ,D_3 } \hfill \\
 {\lambda _1 ,\lambda _2 ,\lambda _3 } \hfill \\
\end{array} }} \right)}\cdot \nabla ^{\left( {{\begin{array}{*{20}c}
 {D_1 ,D_2 ,D_3 } \hfill \\
 {\lambda _1 ,\lambda _2 ,\lambda _3 } \hfill \\
\end{array} }} \right)},
\end{equation}
of the scaling-law scalar field $X$ is defined as
\begin{equation}
\label{eq20}
\begin{array}{l}
 \nabla ^{\left( {{\begin{array}{*{20}c}
 {2D_1 ,2D_2 ,2D_3 } \hfill \\
 {\lambda _1 ,\lambda _2 ,\lambda _3 } \hfill \\
\end{array} }} \right)}X \\
 =\left[ {\left( {\lambda _1 D_1 x^{D_1 -1}} \right)\partial _x^{\left( 1
\right)} } \right]^2X+\left[ {\left( {\lambda _2 D_2 y^{D_2 -1}}
\right)\partial _y^{\left( 1 \right)} } \right]^2X+\left[ {\left( {\lambda
_3 D_3 z^{D_3 -1}} \right)\partial _z^{\left( 1 \right)} } \right]^2X. \\
 \end{array}
\end{equation}
Let the scaling-law vector field, defined by:
\begin{equation}
\label{eq21}
\widehat{{\rm {\bf {\rm O}}}}=\widehat{{\rm {\bf {\rm O}}}}\left( {\lambda
_1 x^{D_1 },\lambda _2 y^{D_2 },\lambda _3 z^{D_3 }} \right)=\widehat{{\rm
O}}_x i+\widehat{{\rm O}}_y j+\widehat{{\rm O}}_z k.
\end{equation}
The scaling-law divergence of the scaling-law vector field $\widehat{{\rm
{\bf {\rm O}}}}$ is defined as
\begin{equation}
\label{eq22}
\begin{array}{l}
\nabla ^{\left( {{\begin{array}{*{20}c}
 {2D_1 ,2D_2 ,2D_3 } \hfill \\
 {\lambda _1 ,\lambda _2 ,\lambda _3 } \hfill \\
\end{array} }} \right)}\cdot \widehat{{\rm {\bf {\rm O}}}}\\
=
\left( {\lambda
_1 D_1 x^{D_1 -1}} \right)\partial _x^{\left( 1 \right)} \widehat{{\rm O}}_x
+\left( {\lambda _2 D_2 y^{D_2 -1}} \right)\partial _y^{\left( 1 \right)}
\widehat{{\rm O}}_y +\left( {\lambda _3 D_3 z^{D_3 -1}} \right)\partial
_z^{\left( 1 \right)} \widehat{{\rm O}}_z .
\end{array}
\end{equation}
The scaling-law curl of the scaling-law vector field $\widehat{{\rm {\bf
{\rm O}}}}$ is defined as
\begin{equation}
\label{eq23}
\begin{array}{l}
 \nabla ^{\left( {{\begin{array}{*{20}c}
 {2D_1 ,2D_2 ,2D_3 } \hfill \\
 {\lambda _1 ,\lambda _2 ,\lambda _3 } \hfill \\
\end{array} }} \right)}\times \widehat{{\rm {\bf {\rm O}}}}\\
=\left(
{{\begin{array}{*{20}c}
 i \hfill & j \hfill & k \hfill \\
 {\left( {\lambda _1 D_1 x^{D_1 -1}} \right)\partial _x^{\left( 1 \right)} }
\hfill & {\left( {\lambda _2 D_2 y^{D_2 -1}} \right)\partial _y^{\left( 1
\right)} } \hfill & {\left( {\lambda _3 D_3 z^{D_3 -1}} \right)\partial
_z^{\left( 1 \right)} } \hfill \\
 {\widehat{{\rm O}}_x } \hfill & {\widehat{{\rm O}}_y } \hfill &
{\widehat{{\rm O}}_z } \hfill \\
\end{array} }} \right) \\
 =\left( {{\begin{array}{*{20}c}
 {\left( {\lambda _2 D_2 y^{D_2 -1}} \right)\partial _y^{\left( 1 \right)} }
\hfill & {\left( {\lambda _3 D_3 z^{D_3 -1}} \right)\partial _z^{\left( 1
\right)} } \hfill \\
 {\widehat{{\rm O}}_y } \hfill & {\widehat{{\rm O}}_z } \hfill \\
\end{array} }} \right)i-\left( {{\begin{array}{*{20}c}
 {\left( {\lambda _1 D_1 x^{D_1 -1}} \right)\partial _x^{\left( 1 \right)} }
\hfill & {\left( {\lambda _3 D_3 z^{D_3 -1}} \right)\partial _z^{\left( 1
\right)} } \hfill \\
 {\widehat{{\rm O}}_x } \hfill & {\widehat{{\rm O}}_z } \hfill \\
\end{array} }} \right)j \\
 +\left( {{\begin{array}{*{20}c}
 {\left( {\lambda _1 D_1 x^{D_1 -1}} \right)\partial _x^{\left( 1 \right)} }
\hfill & {\left( {\lambda _2 D_2 y^{D_2 -1}} \right)\partial _y^{\left( 1
\right)} } \hfill \\
 {\widehat{{\rm O}}_x } \hfill & {\widehat{{\rm O}}_y } \hfill \\
\end{array} }} \right)k \\
 =\left( {\left( {\lambda _2 D_2 y^{D_2 -1}} \right)\partial _y^{\left( 1
\right)} \widehat{{\rm O}}_z -\left( {\lambda _3 D_3 z^{D_3 -1}}
\right)\partial _z^{\left( 1 \right)} \widehat{{\rm O}}_y } \right)i\\
+\left(
{\left( {\lambda _3 D_3 z^{D_3 -1}} \right)\partial _z^{\left( 1 \right)}
\widehat{{\rm O}}_x -\left( {\lambda _1 D_1 x^{D_1 -1}} \right)\partial
_x^{\left( 1 \right)} \widehat{{\rm O}}_z } \right)j \\
 +\left( {\left( {\lambda _1 D_1 x^{D_1 -1}} \right)\partial _x^{\left( 1
\right)} \widehat{{\rm O}}_y -\left( {\lambda _2 D_2 y^{D_2 -1}}
\right)\partial _y^{\left( 1 \right)} \widehat{{\rm O}}_x } \right)k. \\
 \end{array}
\end{equation}
The properties for the scaling-law gradient can be presented as follows:
\begin{equation}
\label{eq24}
\begin{array}{l}
\nabla ^{\left( {{\begin{array}{*{20}c}
 {D_1 ,D_2 ,D_3 } \hfill \\
 {\lambda _1 ,\lambda _2 ,\lambda _3 } \hfill \\
\end{array} }} \right)}\times \left( {\nabla ^{\left(
{{\begin{array}{*{20}c}
 {D_1 ,D_2 ,D_3 } \hfill \\
 {\lambda _1 ,\lambda _2 ,\lambda _3 } \hfill \\
\end{array} }} \right)}\times \widehat{{\rm {\bf {\rm O}}}}} \right)\\
=\nabla
^{\left( {{\begin{array}{*{20}c}
 {D_1 ,D_2 ,D_3 } \hfill \\
 {\lambda _1 ,\lambda _2 ,\lambda _3 } \hfill \\
\end{array} }} \right)}\left( {\nabla ^{\left( {{\begin{array}{*{20}c}
 {D_1 ,D_2 ,D_3 } \hfill \\
 {\lambda _1 ,\lambda _2 ,\lambda _3 } \hfill \\
\end{array} }} \right)}\cdot \widehat{{\rm {\bf {\rm O}}}}} \right)-\nabla
^{\left( {{\begin{array}{*{20}c}
 {2D_1 ,2D_2 ,2D_3 } \hfill \\
 {\lambda _1 ,\lambda _2 ,\lambda _3 } \hfill \\
\end{array} }} \right)}\widehat{{\rm {\bf {\rm O}}}},
\end{array}
\end{equation}

\begin{equation}
\label{eq25}
\nabla ^{\left( {{\begin{array}{*{20}c}
 {D_1 ,D_2 ,D_3 } \hfill \\
 {\lambda _1 ,\lambda _2 ,\lambda _3 } \hfill \\
\end{array} }} \right)}\cdot \left( {\nabla ^{\left( {{\begin{array}{*{20}c}
 {D_1 ,D_2 ,D_3 } \hfill \\
 {\lambda _1 ,\lambda _2 ,\lambda _3 } \hfill \\
\end{array} }} \right)}\times \widehat{{\rm {\bf {\rm O}}}}} \right)=0,
\end{equation}
\begin{equation}
\label{eq26}
\nabla ^{\left( {{\begin{array}{*{20}c}
 {D_1 ,D_2 ,D_3 } \hfill \\
 {\lambda _1 ,\lambda _2 ,\lambda _3 } \hfill \\
\end{array} }} \right)}\times \left( {\nabla ^{\left(
{{\begin{array}{*{20}c}
 {D_1 ,D_2 ,D_3 } \hfill \\
 {\lambda _1 ,\lambda _2 ,\lambda _3 } \hfill \\
\end{array} }} \right)}X} \right)=0,
\end{equation}
and
\begin{equation}
\label{eq27}
\nabla ^{\left( {{\begin{array}{*{20}c}
 {D_1 ,D_2 ,D_3 } \hfill \\
 {\lambda _1 ,\lambda _2 ,\lambda _3 } \hfill \\
\end{array} }} \right)}\left( {XY} \right)=Y\nabla ^{\left(
{{\begin{array}{*{20}c}
 {D_1 ,D_2 ,D_3 } \hfill \\
 {\lambda _1 ,\lambda _2 ,\lambda _3 } \hfill \\
\end{array} }} \right)}X+X\nabla ^{\left( {{\begin{array}{*{20}c}
 {D_1 ,D_2 ,D_3 } \hfill \\
 {\lambda _1 ,\lambda _2 ,\lambda _3 } \hfill \\
\end{array} }} \right)}Y,
\end{equation}
where $Y=Y\left( {\lambda _1 x^{D_1 },\lambda _2 y^{D_2 },\lambda _3 z^{D_3
}} \right)$.

\section{The scaling-law vector calculus }

Let ${\rm {\bf l}}=\left( {\lambda _1 x^{D_1 },\lambda _2 y^{D_2 },\lambda
_3 z^{D_3 }} \right)$ be the scaling-law vector line.

The arc length is presented as follows:
\begin{equation}
\label{eq28}
\begin{array}{l}
\ell
=\int\limits_0^\ell {d\ell } \\
=\int\limits_a^b {\sqrt {\left( {\lambda
_1 D_1 x^{D_1 -1}} \right)^2\left( {\frac{dx}{dt}} \right)^2+\left( {\lambda
_2 D_2 y^{D_2 -1}} \right)^2\left( {\frac{dy}{dt}} \right)^2+\left( {\lambda
_3 D_3 z^{D_3 -1}} \right)^2\left( {\frac{dz}{dt}} \right)^2} dt} ,
\end{array}
\end{equation}
where
\begin{equation}
\label{eq29}
d\ell =\sqrt {\left( {\lambda _1 D_1 x^{D_1 -1}} \right)^2\left(
{\frac{dx}{dt}} \right)^2+\left( {\lambda _2 D_2 y^{D_2 -1}} \right)^2\left(
{\frac{dy}{dt}} \right)^2+\left( {\lambda _3 D_3 z^{D_3 -1}} \right)^2\left(
{\frac{dz}{dt}} \right)^2} dt.
\end{equation}
The scaling-law line integral of the scaling-law vector field $\widehat{{\rm
{\bf {\rm O}}}}$ along the curve ${\rm {\bf l}}$, denoted by $\Pi $, is
defined as
\begin{equation}
\label{eq30}
\Pi =\int\limits_\ell {\widehat{{\rm {\bf {\rm O}}}}\cdot d{\rm {\bf l}}}
=\int\limits_\ell {\widehat{{\rm {\bf {\rm O}}}}\cdot {\rm {\bf n}}d\ell }
,
\end{equation}
which leads to
\begin{equation}
\label{eq31}
\begin{array}{l}
 \Pi =\int\limits_\ell {\widehat{{\rm {\bf {\rm O}}}}\cdot d{\rm {\bf l}}}
=\int\limits_\ell {\widehat{{\rm {\bf {\rm O}}}}\cdot {\rm {\bf n}}d\ell }
\\
 =\int\limits_\ell {\left( {\lambda _1 D_1 x^{D_1 -1}} \right)\widehat{{\rm
O}}_x dx+\left( {\lambda _2 D_2 y^{D_2 -1}} \right)\widehat{{\rm O}}_y
dy+\left( {\lambda _3 D_3 z^{D_3 -1}} \right)\widehat{{\rm O}}_z dz} , \\
 \end{array}
\end{equation}
where the element of the scaling-law line is
\begin{equation}
\label{eq32}
\begin{array}{l}
 d{\rm {\bf l}}={\rm {\bf n}}d\ell \\
 =i\left( {\lambda _1 D_1 x^{D_1 -1}}
\right)dx+j\left( {\lambda _2 D_2 y^{D_2 -1}} \right)dy+k\left( {\lambda _3
D_3 z^{D_3 -1}} \right)dz \\
 =id\left( {\lambda _1 x^{D_1 }} \right)+jd\left( {\lambda _2 y^{D_2 }}
\right)+kd\left( {\lambda _3 z^{D_3 }} \right) \\
 \end{array}
\end{equation}
with the unit vector ${\rm {\bf n}}$ tangent to the scaling-law vector line
${\rm {\bf l}}$.

From (\ref{eq31}) we give
\begin{equation}
\label{eq33}
\Pi =\int\limits_a^b {\left[ {\left( {\lambda _1 D_1 x^{D_1 -1}}
\right)\widehat{{\rm O}}_x \frac{dx}{dt}+\left( {\lambda _2 D_2 y^{D_2 -1}}
\right)\widehat{{\rm O}}_y \frac{dy}{dt}+\left( {\lambda _3 D_3 z^{D_3 -1}}
\right)\widehat{{\rm O}}_z \frac{dz}{dt}} \right]dt}
\end{equation}
since
\begin{equation}
\label{eq34}
\begin{array}{l}
 \left( {\lambda _1 D_1 x^{D_1 -1}} \right)\widehat{{\rm O}}_x dx+\left(
{\lambda _2 D_2 y^{D_2 -1}} \right)\widehat{{\rm O}}_y dy+\left( {\lambda _3
D_3 z^{D_3 -1}} \right)\widehat{{\rm O}}_z dz \\
 =\left[ {\left( {\lambda _1 D_1 x^{D_1 -1}} \right)\widehat{{\rm O}}_x
\frac{dx}{dt}+\left( {\lambda _2 D_2 y^{D_2 -1}} \right)\widehat{{\rm O}}_y
\frac{dy}{dt}+\left( {\lambda _3 D_3 z^{D_3 -1}} \right)\widehat{{\rm O}}_z
\frac{dz}{dt}} \right]dt. \\
 \end{array}
\end{equation}
Let $S=S\left( {\lambda _1 x^{D_1 },\lambda _2 y^{D_2 }} \right)$.

The scaling-law double integral of the scaling-law scalar field $X$ on the
region $S$, denoted by${\rm M}\left( X \right)$, is defined as
\begin{equation}
\label{eq35}
\begin{array}{l}
{\rm M}\left( X \right)=\int\!\!\!\int\limits_S {XdS}\\
=\int\!\!\!\int\limits_S {X\left( {\lambda _1 D_1 x^{D_1 -1}} \right)\left(
{\lambda _2 D_2 y^{D_2 -1}} \right)dxdy} \\
=\int\!\!\!\int\limits_S {{\rm
T}d\left( {\lambda _1 x^{D_1 }} \right)d\left( {\lambda _2 y^{D_2 }}
\right)} ,
\end{array}
\end{equation}
where $dS=\left( {\lambda _1 D_1 x^{D_1 -1}} \right)\left( {\lambda _2 D_2
y^{D_2 -1}} \right)dxdy=d\left( {\lambda _1 x^{D_1 }} \right)d\left(
{\lambda _2 y^{D_2 }} \right)$ is the element of the scaling-law area.

Thus, we have that
\begin{equation}
\label{eq36}
\begin{array}{l}
 {\rm M}\left( X \right)=\int\!\!\!\int\limits_S {XdS} \\
 =\int\limits_c^d {\left[ {\int\limits_a^b {X\left( {\lambda _1 D_1 x^{D_1
-1}} \right)dx} } \right]\left( {\lambda _2 D_2 y^{D_2 -1}} \right)dy}\\
=\int\limits_a^b {\left[ {\int\limits_c^d {X\left( {\lambda _2 D_2 y^{D_2
-1}} \right)dy} } \right]\left( {\lambda _1 D_1 x^{D_1 -1}} \right)dx} \\
 =\int\limits_c^d {\left[ {\int\limits_a^b {Xd\left( {\lambda _1 x^{D_1 }}
\right)} } \right]d\left( {\lambda _2 y^{D_2 }} \right)} =\int\limits_a^b
{\left[ {\int\limits_c^d {Xd\left( {\lambda _2 y^{D_2 }} \right)} }
\right]d\left( {\lambda _1 x^{D_1 }} \right)} , \\
 \end{array}
\end{equation}
where $x\in \left[ {a,b} \right]$ and $y\in \left[ {c,d} \right]$.

The scaling-law volume integral of the scaling-law scalar field $X$ in the
domain $\Omega $ is defined as
\begin{equation}
\label{eq37}
\begin{array}{l}
 V\left( X \right)=\mathop{\int\!\!\!\int\!\!\!\int}\limits_{\kern-5.5pt
\Omega } {XdV} \\
 =\mathop{\int\!\!\!\int\!\!\!\int}\limits_{\kern-5.5pt \Omega } {X\left(
{\lambda _1 D_1 x^{D_1 -1}} \right)\left( {\lambda _2 D_2 y^{D_2 -1}}
\right)\left( {\lambda _3 D_3 z^{D_3 -1}} \right)dxdydz} \\
 =\mathop{\int\!\!\!\int\!\!\!\int}\limits_{\kern-5.5pt \Omega } {Xd\left(
{\lambda _1 x^{D_1 }} \right)d\left( {\lambda _2 y^{D_2 }} \right)d\left(
{\lambda _3 z^{D_3 }} \right)} , \\
 \end{array}
\end{equation}
where
\[
\begin{array}{l}
 dV=\left( {\lambda _1 D_1 x^{D_1 -1}} \right)\left( {\lambda _2 D_2 y^{D_2
-1}} \right)\left( {\lambda _3 D_3 z^{D_3 -1}} \right)dxdydz \\
 { }=d\left( {\lambda _1 x^{D_1 }} \right)d\left( {\lambda _2 y^{D_2 }}
\right)d\left( {\lambda _3 z^{D_3 }} \right) \\
 \end{array}
\]
is the element of volume.

Thus, we have that
\begin{equation}
\label{eq38}
\begin{array}{l}
 \mathop{\int\!\!\!\int\!\!\!\int}\limits_{\kern-5.5pt \Omega } {XdV}
=\int\limits_f^g {\left[ {\int\limits_c^d {\left( {\int\limits_a^b {X\left(
{\lambda _1 D_1 x^{D_1 -1}} \right)dx} } \right)\left( {\lambda _2 D_2
y^{D_2 -1}} \right)dy} } \right]\left( {\lambda _3 D_3 z^{D_3 -1}}
\right)dz} \\
 =\int\limits_f^g {\left[ {\int\limits_a^b {\left( {\int\limits_c^d {X\left(
{\lambda _2 D_2 y^{D_2 -1}} \right)dy} } \right)\left( {\lambda _1 D_1
x^{D_1 -1}} \right)dx} } \right]\left( {\lambda _3 D_3 z^{D_3 -1}}
\right)dz} \\
 =\int\limits_a^b {\left[ {\int\limits_c^d {\left( {\int\limits_e^f {X\left(
{\lambda _3 D_3 z^{D_3 -1}} \right)dz} } \right)\left( {\lambda _2 D_2
y^{D_2 -1}} \right)dy} } \right]\left( {\lambda _1 D_1 x^{D_1 -1}}
\right)dx} \\
 =\int\limits_f^g {\left[ {\int\limits_c^d {\left( {\int\limits_a^b
{Xd\left( {\lambda _1 x^{D_1 }} \right)} } \right)d\left( {\lambda _2 y^{D_2
}} \right)} } \right]d\left( {\lambda _3 z^{D_3 }} \right)} \\
 =\int\limits_f^g {\left[ {\int\limits_a^b {\left( {\int\limits_c^d
{Xd\left( {\lambda _2 y^{D_2 }} \right)} } \right)d\left( {\lambda _1 x^{D_1
}} \right)} } \right]d\left( {\lambda _3 z^{D_3 }} \right)} \\
 =\int\limits_a^b {\left[ {\int\limits_c^d {\left( {\int\limits_e^f
{Xd\left( {\lambda _3 z^{D_3 }} \right)} } \right)d\left( {\lambda _2 y^{D_2
}} \right)} } \right]d\left( {\lambda _1 x^{D_1 }} \right)} , \\
 \end{array}
\end{equation}
where $x\in \left[ {a,b} \right]$, $y\in \left[ {c,d} \right]$ and $z\in
\left[ {f,g} \right]$.

Let the scaling-law surface be defined by ${\rm {\bf S}}={\rm {\bf S}}\left(
{\lambda _1 x^{D_1 },\lambda _2 y^{D_2 },\lambda _3 z^{D_3 }} \right)$.

The scaling-law surface integral of the scaling-law vector field
$\widehat{{\rm {\bf {\rm O}}}}$ on the scaling-law surface $\partial \Omega
$ of the domain $\Omega $ is defined as
\begin{equation}
\label{eq39}
\int\!\!\!\int\limits_{\partial \Omega } {\widehat{{\rm {\bf {\rm O}}}}\cdot
{\rm {\bf dS}}} =\int\!\!\!\int\limits_{\partial \Omega } {\widehat{{\rm
{\bf {\rm O}}}}\cdot {\rm {\bf a}}dS} ,
\end{equation}
where ${\rm {\bf a}}={\rm {\bf dS}}/\left| {{\rm {\bf dS}}} \right|={\rm
{\bf dS}}/dS$ is the unit normal vector to the scaling-law surface $\partial
\Omega $ with $dS=\left| {{\rm {\bf dS}}} \right|$, and
\begin{equation}
\label{eq40}
\begin{array}{l}
 {\rm {\bf dS}}=id\left( {\lambda _2 y^{D_2 }} \right)d\left( {\lambda _3
z^{D_3 }} \right)+jd\left( {\lambda _1 x^{D_1 }} \right)d\left( {\lambda _3
z^{D_3 }} \right)+kd\left( {\lambda _1 x^{D_1 }} \right)d\left( {\lambda _2
y^{D_2 }} \right) \\
{ }=i\left( {\lambda _2 D_2 y^{D_2 -1}} \right)\left( {\lambda _3 D_3
z^{D_3 -1}} \right)dydz+j\left( {\lambda _1 D_1 x^{D_1 -1}} \right)\left(
{\lambda _3 D_3 z^{D_3 -1}} \right)dxdz \\
 +k\left( {\lambda _1 D_1 x^{D_1 -1}} \right)\left( {\lambda _2 D_2 y^{D_2
-1}} \right)dxdy \\
 \end{array}
\end{equation}
is the element of the scaling-law surface.

From (\ref{eq39}) and (\ref{eq40}) we present
\begin{equation}
\label{eq41}
\begin{array}{l}
 \int\!\!\!\int\limits_{\partial \Omega } {\widehat{{\rm {\bf {\rm
O}}}}\cdot {\rm {\bf dS}}} \\
 =\int\!\!\!\int\limits_{\partial \Omega } {\widehat{{\rm O}}_x d\left(
{\lambda _2 y^{D_2 }} \right)d\left( {\lambda _3 z^{D_3 }}
\right)+\widehat{{\rm O}}_y d\left( {\lambda _1 x^{D_1 }} \right)d\left(
{\lambda _3 z^{D_3 }} \right)+\widehat{{\rm O}}_z d\left( {\lambda _1 x^{D_1
}} \right)d\left( {\lambda _2 y^{D_2 }} \right)} \\
 =\int\!\!\!\int\limits_{\partial \Omega } {\widehat{{\rm O}}_x \left(
{\lambda _2 D_2 y^{D_2 -1}} \right)\left( {\lambda _3 D_3 z^{D_3 -1}}
\right)dydz} +\int\!\!\!\int\limits_{\partial \Omega } {\widehat{{\rm O}}_y
\left( {\lambda _1 D_1 x^{D_1 -1}} \right)\left( {\lambda _3 D_3 z^{D_3 -1}}
\right)dxdz} \\
 +\int\!\!\!\int\limits_{\partial \Omega } {\widehat{{\rm O}}_z \left(
{\lambda _1 D_1 x^{D_1 -1}} \right)\left( {\lambda _2 D_2 y^{D_2 -1}}
\right)dxdy} . \\
 \end{array}
\end{equation}
The flux of the scaling-law vector field $\widehat{{\rm {\bf {\rm O}}}}$
across the scaling-law surface $\partial \Omega $, denoted by $G\left(
{\widehat{{\rm {\bf {\rm O}}}}} \right)$, is defined as
\begin{equation}
\label{eq42}
G\left( {\widehat{{\rm {\bf {\rm O}}}}}
\right)=\mathop{{\int\!\!\!\!\!\int}\mkern-21mu \bigcirc}\limits_{\partial
\Omega } {\widehat{{\rm {\bf {\rm O}}}}\cdot {\rm {\bf dS}}} .
\end{equation}
The scaling-law divergence of the scaling-law vector field $\widehat{{\rm
{\bf {\rm O}}}}$ is defined as
\begin{equation}
\label{eq43}
\nabla ^{\left( {{\begin{array}{*{20}c}
 {D_1 ,D_2 ,D_3 } \hfill \\
 {\lambda _1 ,\lambda _2 ,\lambda _3 } \hfill \\
\end{array} }} \right)}\cdot \widehat{{\rm {\bf {\rm O}}}}=\mathop {\lim
}\limits_{\Delta V_m \to 0} \frac{1}{\Delta V_m
}\mathop{{\int\!\!\!\!\!\int}\mkern-21mu \bigcirc}\limits_{\Delta \partial
\Omega _m } {\widehat{{\rm {\bf {\rm O}}}}\cdot {\rm {\bf dS}}} ,
\end{equation}
where the scaling-law volume $V$ is divided into a large number of small
subvolumes $\Delta V_m $ with the scaling-law surfaces $\Delta \Omega _m $,
and ${\rm {\bf dS}}$ is the element of the scaling-law surface $\partial
\Omega $ bounding the solid $\Omega $.

Here, (\ref{eq14}) is equal to (\ref{eq43}) in the Cartesian-type coordinate system.

The scaling-law curl of the scaling-law vector field $\widehat{{\rm {\bf
{\rm O}}}}$ is defined as
\begin{equation}
\label{eq44}
\nabla ^{\left( {{\begin{array}{*{20}c}
 {D_1 ,D_2 ,D_3 } \hfill \\
 {\lambda _1 ,\lambda _2 ,\lambda _3 } \hfill \\
\end{array} }} \right)}\times \widehat{{\rm {\bf {\rm O}}}}=\mathop {\lim
}\limits_{\Delta S_m \to 0} \frac{1}{\Delta S_m }\oint\limits_{\Delta \ell
_m } {\widehat{{\rm {\bf {\rm O}}}}\cdot d{\rm {\bf l}}} ,
\end{equation}
where $d{\rm {\bf l}}$ is the element of the scaling-law vector line,
$\Delta S_m $ is a small scaling-law surface element perpendicular to ${\rm
{\bf n}}$, $\Delta \ell _m $ is the closed curve of the boundary of $\Delta
S_m $, and ${\rm {\bf n}}$ is oriented in a positive sense.

Here, (\ref{eq15}) is (\ref{eq44}) in the Cartesian-type coordinate system.

From (\ref{eq43}) we present the Gauss-like theorem for the scaling-law vector
calculus as follows.

Let us consider that
\begin{equation}
\label{eq45}
\begin{array}{l}
 \mathop{{\int\!\!\!\!\!\int}\mkern-21mu \bigcirc}\limits_{\partial \Omega }
{\widehat{{\rm {\bf {\rm O}}}}\cdot {\rm {\bf n}}dS}
=\mathop{{\int\!\!\!\!\!\int}\mkern-21mu \bigcirc}\limits_{\partial \Omega }
{\widehat{{\rm {\bf {\rm O}}}}\cdot {\rm {\bf dS}}} \\
{ }=\mathop{{\int\!\!\!\!\!\int}\mkern-21mu \bigcirc}\limits_{\partial
\Omega } {\widehat{{\rm O}}_x d\left( {\lambda _2 y^{D_2 }} \right)d\left(
{\lambda _3 z^{D_3 }} \right)+\widehat{{\rm O}}_y d\left( {\lambda _1 x^{D_1
}} \right)d\left( {\lambda _3 z^{D_3 }} \right)+\widehat{{\rm O}}_z d\left(
{\lambda _1 x^{D_1 }} \right)d\left( {\lambda _2 y^{D_2 }} \right)} \\
{ }=\mathop{{\int\!\!\!\!\!\int}\mkern-21mu \bigcirc}\limits_{\partial
\Omega } {\widehat{{\rm O}}_x \left( {\lambda _2 D_2 y^{D_2 -1}}
\right)\left( {\lambda _3 D_3 z^{D_3 -1}} \right)dydz}
+\mathop{{\int\!\!\!\!\!\int}\mkern-21mu \bigcirc}\limits_{\partial \Omega }
{\widehat{{\rm O}}_y \left( {\lambda _1 D_1 x^{D_1 -1}} \right)\left(
{\lambda _3 D_3 z^{D_3 -1}} \right)dxdz} \\
 +\mathop{{\int\!\!\!\!\!\int}\mkern-21mu \bigcirc}\limits_{\partial \Omega
} {\widehat{{\rm O}}_z \left( {\lambda _1 D_1 x^{D_1 -1}} \right)\left(
{\lambda _2 D_2 y^{D_2 -1}} \right)dxdy} . \\
 \end{array}
\end{equation}
The Gauss-Ostrogradsky-like theorem for the scaling-law vector calculus
states that
\begin{equation}
\label{eq46}
\mathop{\int\!\!\!\int\!\!\!\int}\limits_{\kern-5.5pt \Omega } {\nabla
^{\left( {{\begin{array}{*{20}c}
 {D_1 ,D_2 ,D_3 } \hfill \\
 {\lambda _1 ,\lambda _2 ,\lambda _3 } \hfill \\
\end{array} }} \right)}\cdot \widehat{{\rm {\bf {\rm O}}}}dV}
=\mathop{{\int\!\!\!\!\!\int}\mkern-21mu \bigcirc}\limits_{\partial \Omega }
{\widehat{{\rm {\bf {\rm O}}}}\cdot {\rm {\bf a}}dS}
\end{equation}
or
\begin{equation}
\label{eq47}
\mathop{\int\!\!\!\int\!\!\!\int}\limits_{\kern-5.5pt \Omega } {\nabla
^{\left( {{\begin{array}{*{20}c}
 {D_1 ,D_2 ,D_3 } \hfill \\
 {\lambda _1 ,\lambda _2 ,\lambda _3 } \hfill \\
\end{array} }} \right)}\cdot \widehat{{\rm {\bf {\rm O}}}}dV}
=\mathop{{\int\!\!\!\!\!\int}\mkern-21mu \bigcirc}\limits_{\partial \Omega }
{\widehat{{\rm {\bf {\rm O}}}}\cdot {\rm {\bf dS}}} .
\end{equation}
When $D_1 =D_2 =D_3 =1$ and $\lambda _1 =\lambda _2 =\lambda _3 =1$, (\ref{eq46})
becomes the Gauss-Ostrogradsky theorem, proposed by Gauss in 1813 \cite{27} and
by Ostrogradsky in 1828 \cite{28}.

From (\ref{eq44}) we present the Stokes-like theorem for the scaling-law vector
calculus as follows.

We now consider that
\begin{equation}
\label{eq48}
\begin{array}{l}
 \oint\limits_\ell {\widehat{{\rm {\bf {\rm O}}}}\cdot {\rm {\bf dl}}}\\
=\oint\limits_\ell {\widehat{{\rm O}}_x \left( {\lambda _1 D_1 x^{D_1 -1}}
\right)dx+\widehat{{\rm O}}_y \left( {\lambda _2 D_2 y^{D_2 -1}}
\right)dy+\widehat{{\rm O}}_z \left( {\lambda _3 D_3 z^{D_3 -1}} \right)dz}
\\
 =\oint\limits_\ell {\widehat{{\rm O}}_x d\left( {\lambda _1 x^{D_1 }}
\right)+\widehat{{\rm O}}_y d\left( {\lambda _2 y^{D_2 }}
\right)+\widehat{{\rm O}}_z d\left( {\lambda _3 z^{D_3 }} \right)} . \\
 \end{array}
\end{equation}
The Stokes-like theorem for the scaling-law vector calculus states that
\begin{equation}
\label{eq49}
\mathop{{\int\!\!\!\!\!\int}\mkern-21mu \bigcirc}\limits_{\partial \Omega }
{\left( {\nabla ^{\left( {{\begin{array}{*{20}c}
 {D_1 ,D_2 ,D_3 } \hfill \\
 {\lambda _1 ,\lambda _2 ,\lambda _3 } \hfill \\
\end{array} }} \right)}\times \widehat{{\rm {\bf {\rm O}}}}} \right)\cdot
{\rm {\bf a}}dS} =\oint\limits_\ell {\widehat{{\rm {\bf {\rm O}}}}\cdot {\rm
{\bf dl}}}
\end{equation}
or
\begin{equation}
\label{eq50}
\mathop{{\int\!\!\!\!\!\int}\mkern-21mu \bigcirc}\limits_{\partial \Omega }
{\left( {\nabla ^{\left( {{\begin{array}{*{20}c}
 {D_1 ,D_2 ,D_3 } \hfill \\
 {\lambda _1 ,\lambda _2 ,\lambda _3 } \hfill \\
\end{array} }} \right)}\times \widehat{{\rm {\bf {\rm O}}}}} \right)\cdot
{\rm {\bf dS}}} =\oint\limits_\ell {\widehat{{\rm {\bf {\rm O}}}}\cdot {\rm
{\bf dl}}} .
\end{equation}
Here, when $D_1 =D_2 =D_3 =1$ and $\lambda _1 =\lambda _2 =\lambda _3 =1$,
(\ref{eq49}) is the Stokes theorem, proposed by Stokes in 1845 \cite{29}.

With use of (\ref{eq49}) and (\ref{eq50}), we present the Green-like theorem for the
scaling-law vector calculus as follows.

The Green-like theorem for the scaling-law vector calculus states
\begin{equation}
\label{eq51}
\begin{array}{l}
 \oint\limits_\ell {\left( {\lambda _1 D_1 x^{D_1 -1}} \right)\widehat{{\rm
O}}_x dx+\left( {\lambda _2 D_2 y^{D_2 -1}} \right)\widehat{{\rm O}}_y dy}
\\
 =\int\!\!\!\int\limits_S {\left( {\left( {\lambda _1 D_1 x^{D_1 -1}}
\right){ }^{SL}\partial _x^{\left( 1 \right)} \widehat{{\rm O}}_y -\left(
{\lambda _2 D_2 y^{D_2 -1}} \right){ }^{SL}\partial _y^{\left( 1 \right)}
\widehat{{\rm O}}_x } \right)\left( {\lambda _1 D_1 x^{D_1 -1}}
\right)\left( {\lambda _2 D_2 y^{D_2 -1}} \right)dxdy} , \\
 \end{array}
\end{equation}
where $S$ is the domain bounded by the scaling-law contour $\ell $.

The Green-like identity of first type via scaling-law vector calculus states
that
\begin{equation}
\label{eq52}
\begin{array}{l}
\mathop{\int\!\!\!\int\!\!\!\int}\limits_{\kern-5.5pt \Omega } {\nabla
^{\left( {{\begin{array}{*{20}c}
 {D_1 ,D_2 ,D_3 } \hfill \\
 {\lambda _1 ,\lambda _2 ,\lambda _3 } \hfill \\
\end{array} }} \right)}\cdot \left( {X\nabla ^{\left(
{{\begin{array}{*{20}c}
 {2D_1 ,2D_2 ,2D_3 } \hfill \\
 {\lambda _1 ,\lambda _2 ,\lambda _3 } \hfill \\
\end{array} }} \right)}Y+\nabla ^{\left( {{\begin{array}{*{20}c}
 {D_1 ,D_2 ,D_3 } \hfill \\
 {\lambda _1 ,\lambda _2 ,\lambda _3 } \hfill \\
\end{array} }} \right)}Y\cdot \nabla ^{\left( {{\begin{array}{*{20}c}
 {D_1 ,D_2 ,D_3 } \hfill \\
 {\lambda _1 ,\lambda _2 ,\lambda _3 } \hfill \\
\end{array} }} \right)}X} \right)dV}\\
=\mathop{{\int\!\!\!\!\!\int}\mkern-21mu \bigcirc}\limits_{\partial \Omega }
{X\partial _u^{\left( {{\begin{array}{*{20}c}
 {D_1 ,D_2 ,D_3 } \hfill \\
 {\lambda _1 ,\lambda _2 ,\lambda _3 } \hfill \\
\end{array} }} \right)} YdS} .
\end{array}
\end{equation}
The Green-like identity of second type via scaling-law vector calculus
states that
\begin{equation}
\label{eq53}
\begin{array}{l}
\mathop{\int\!\!\!\int\!\!\!\int}\limits_{\kern-5.5pt \Omega } {\nabla
^{\left( {{\begin{array}{*{20}c}
 {D_1 ,D_2 ,D_3 } \hfill \\
 {\lambda _1 ,\lambda _2 ,\lambda _3 } \hfill \\
\end{array} }} \right)}\cdot \left( {X\nabla ^{\left(
{{\begin{array}{*{20}c}
 {2D_1 ,2D_2 ,2D_3 } \hfill \\
 {\lambda _1 ,\lambda _2 ,\lambda _3 } \hfill \\
\end{array} }} \right)}Y-Y\nabla ^{\left( {{\begin{array}{*{20}c}
 {2D_1 ,2D_2 ,2D_3 } \hfill \\
 {\lambda _1 ,\lambda _2 ,\lambda _3 } \hfill \\
\end{array} }} \right)}X} \right)dV}\\
=\mathop{{\int\!\!\!\!\!\int}\mkern-21mu \bigcirc}\limits_{\partial \Omega }
{\left( {X\partial _u^{\left( {{\begin{array}{*{20}c}
 {D_1 ,D_2 ,D_3 } \hfill \\
 {\lambda _1 ,\lambda _2 ,\lambda _3 } \hfill \\
\end{array} }} \right)} Y-Y\partial _u^{\left( {{\begin{array}{*{20}c}
 {D_1 ,D_2 ,D_3 } \hfill \\
 {\lambda _1 ,\lambda _2 ,\lambda _3 } \hfill \\
\end{array} }} \right)} X} \right)dS} .
\end{array}
\end{equation}
Here, the Green theorem and identities, proposed by Green in 1828 \cite{30}, are
the special cases of the Green-like theorem and identities when $D_1 =D_2
=D_3 =1$ and $\lambda _1 =\lambda _2 =\lambda _3 =1$.

\section{On the Navier-Stokes-type equations of the scaling-law flow}

Let us consider the coordinate system, defined as
\begin{equation}
\label{eq54}
\left( {\lambda _0 t^{D_0 },\lambda _1 x^{D_1 },\lambda _2 y^{D_2 },\lambda
_3 z^{D_3 }} \right)=\lambda _0 t^{D_0 }+i\lambda _1 x^{D_1 }+j\lambda _2
y^{D_2 }+k\lambda _3 z^{D_3 },
\end{equation}
where $i$, $j$ and $k$ are the unit vector, and $\lambda _0 t^{D_0 }$ is the
fractal scaling law \cite{31} with the normalization constant $\lambda _0 \ge 0$,
the time $t\ge 0$, and the scaling exponent $-\infty <D_0 <\infty $.

Let $\Xi =\Xi \left( {\lambda _0 t^{D_0 },\lambda _1 x^{D_1 },\lambda _2
y^{D_2 },\lambda _3 z^{D_3 }} \right)$ be the scaling-law scalar fluid
field.

The total scaling-law differential of the scaling-law scalar field is given
as follows:
\begin{equation}
\label{eq55}
\begin{array}{l}
 d\Xi \\
 =\left( {\lambda _0 D_0 t^{D_0 -1}} \right)\partial _t^{\left( 1
\right)} \Xi dt+\left( {\lambda _1 D_1 x^{D_1 -1}} \right)\partial
_x^{\left( 1 \right)} \Xi dx+\left( {\lambda _2 D_2 y^{D_2 -1}}
\right)\partial _y^{\left( 1 \right)} \Xi dy \\
 +\left( {\lambda _3 D_3 z^{D_3 -1}} \right)\partial _z^{\left( 1 \right)}
\Xi dz, \\
 \end{array}
\end{equation}
which leads to
\begin{equation}
\label{eq56}
\begin{array}{l}
 \frac{d\Xi }{dt}\\
 =\left( {\lambda _0 D_0 t^{D_0 -1}} \right)\partial
_t^{\left( 1 \right)} \Xi +\left( {\lambda _1 D_1 x^{D_1 -1}}
\right)\partial _x^{\left( 1 \right)} \Xi \frac{dx}{dt}+\left( {\lambda _2
D_2 y^{D_2 -1}} \right)\partial _y^{\left( 1 \right)} \Xi \frac{dy}{dt} \\
 +\left( {\lambda _3 D_3 z^{D_3 -1}} \right)\partial _z^{\left( 1 \right)}
\Xi \frac{dz}{dt}. \\
 \end{array}
\end{equation}
From (\ref{eq56}) the material scaling-law derivative of the scaling-law fluid
density $\Xi $ is defined as follows:
\begin{equation}
\label{eq57}
\frac{D\Xi }{Dt}=\left( {\lambda _0 D_0 t^{D_0 -1}} \right)\partial
_t^{\left( 1 \right)} \Xi +{\rm {\bf \upsilon }}\cdot \nabla ^{\left(
{{\begin{array}{*{20}c}
 {D_1 ,D_2 ,D_3 } \hfill \\
 {\lambda _1 ,\lambda _2 ,\lambda _3 } \hfill \\
\end{array} }} \right)}\Xi ,
\end{equation}
where ${\rm {\bf \upsilon }}=\left( {\partial x/\partial t,\partial
y/\partial t,\partial z/\partial t} \right)=i\upsilon _x +j\upsilon _y
+k\upsilon _z $ are denoted as the velocity vector.

When $D_0 =D_1 =D_2 =D_3 =1$ and $\lambda _0 =\lambda _1 =\lambda _2
=\lambda _3 =1$, (\ref{eq55}) is the Euler notation of the material derivative \cite{32},
and (\ref{eq57}) is the Stokes notation of the material derivative \cite{33,34}.

From (\ref{eq57}) the transport theorem for the scaling-law flow can be given as
follows:
\begin{equation}
\label{eq58}
\frac{D}{Dt}\mathop{\int\!\!\!\int\!\!\!\int}\limits_{\kern-5.5pt {\Omega
\left( t \right)}} {\Xi dV}
=\mathop{\int\!\!\!\int\!\!\!\int}\limits_{\kern-5.5pt {\Omega \left( t
\right)}} {\left[ {\left( {\lambda _0 D_0 t^{D_0 -1}} \right)\partial
_t^{\left( 1 \right)} \Xi +{\rm {\bf \upsilon }}\cdot \nabla ^{\left(
{{\begin{array}{*{20}c}
 {D_1 ,D_2 ,D_3 } \hfill \\
 {\lambda _1 ,\lambda _2 ,\lambda _3 } \hfill \\
\end{array} }} \right)}\Xi } \right]dV},
\end{equation}
which, by using (\ref{eq46}), yields that
\begin{equation}
\label{eq59}
\frac{D}{Dt}\mathop{\int\!\!\!\int\!\!\!\int}\limits_{\kern-5.5pt {\Omega
\left( t \right)}} {\Xi dV}
=\mathop{\int\!\!\!\int\!\!\!\int}\limits_{\kern-5.5pt {\Omega \left( t
\right)}} {\left( {\lambda _0 D_0 t^{D_0 -1}} \right)\partial _t^{\left( 1
\right)} \Xi dV} +\mathop{{\int\!\!\!\!\!\int}\mkern-21mu
\bigcirc}\limits_{\partial \Omega \left( t \right)} {\Xi {\rm {\bf \upsilon
}}\cdot {\rm {\bf dS}}}
\end{equation}
since
\begin{equation}
\label{eq60}
\mathop{\int\!\!\!\int\!\!\!\int}\limits_{\kern-5.5pt {\Omega \left( t
\right)}} {{\rm {\bf \upsilon }}\cdot \nabla ^{\left(
{{\begin{array}{*{20}c}
 {D_1 ,D_2 ,D_3 } \hfill \\
 {\lambda _1 ,\lambda _2 ,\lambda _3 } \hfill \\
\end{array} }} \right)}\Xi dV} =\mathop{{\int\!\!\!\!\!\int}\mkern-21mu
\bigcirc}\limits_{\partial \Omega \left( t \right)} {\Xi \left( {{\rm {\bf
\upsilon }}\cdot {\rm {\bf a}}} \right)dS}
=\mathop{{\int\!\!\!\!\!\int}\mkern-21mu \bigcirc}\limits_{\partial \Omega
\left( t \right)} {\Xi {\rm {\bf \upsilon }}\cdot {\rm {\bf dS}}} ,
\end{equation}
where $\partial \Omega \left( t \right)$ is the surface of $\Omega \left( t
\right)$, ${\rm {\bf a}}$ is the unit normal to the scaling-law
surface, and ${\rm {\bf \upsilon }}$ is the velocity vector.

Taking $D_0 =D_1 =D_2 =D_3 =1$ and $\lambda _0 =\lambda _1 =\lambda _2
=\lambda _3 =1$, we obtain the Reynolds transport theorem \cite{35}.

The conservation of the mass of the scaling-law flow is given as
\begin{equation}
\label{eq61}
\left( {\lambda _0 D_0 t^{D_0 -1}} \right)\partial _t^{\left( 1 \right)}
\rho +{\rm {\bf \upsilon }}\cdot \nabla ^{\left( {{\begin{array}{*{20}c}
 {D_1 ,D_2 ,D_3 } \hfill \\
 {\lambda _1 ,\lambda _2 ,\lambda _3 } \hfill \\
\end{array} }} \right)}\rho =0
\end{equation}
or
\begin{equation}
\label{eq62}
\left( {\lambda _0 D_0 t^{D_0 -1}} \right)\partial _t^{\left( 1 \right)}
\rho +\nabla ^{\left( {{\begin{array}{*{20}c}
 {D_1 ,D_2 ,D_3 } \hfill \\
 {\lambda _1 ,\lambda _2 ,\lambda _3 } \hfill \\
\end{array} }} \right)}\cdot \left( {{\rm {\bf \upsilon }}\rho } \right)=0
\end{equation}
because
\begin{equation}
\label{eq63}
\frac{D}{Dt}\mathop{\int\!\!\!\int\!\!\!\int}\limits_{\kern-5.5pt {\Omega
\left( t \right)}} {\rho dV}
=\mathop{\int\!\!\!\int\!\!\!\int}\limits_{\kern-5.5pt {\Omega \left( t
\right)}} {\left[ {\left( {\lambda _0 D_0 t^{D_0 -1}} \right)\partial
_t^{\left( 1 \right)} \rho +{\rm {\bf \upsilon }}\cdot \nabla ^{\left(
{{\begin{array}{*{20}c}
 {D_1 ,D_2 ,D_3 } \hfill \\
 {\lambda _1 ,\lambda _2 ,\lambda _3 } \hfill \\
\end{array} }} \right)}\rho } \right]dV} ,
\end{equation}
which is derived from the mass of the scaling-law flow, defined as
\begin{equation}
\label{eq64}
{\rm M}=\mathop{\int\!\!\!\int\!\!\!\int}\limits_{\kern-5.5pt {\Omega \left(
t \right)}} {\rho dV}
\end{equation}
where $\rho $ and ${\rm M}$ are the density and mass of the scaling-law
flow, respectively.

Here, for $D_0 =D_1 =D_2 =D_3 =1$ and $\lambda _0 =\lambda _1 =\lambda _2
=\lambda _3 =1$, (\ref{eq61}) is the conservation of the mass \cite{32}.

The Cauchy-type strain tensor for the scaling-law flow, denoted by $\varpi
$, is defined as
\begin{equation}
\label{eq65}
\varpi =\frac{1}{2}\left( {\nabla ^{\left( {{\begin{array}{*{20}c}
 {D_1 ,D_2 ,D_3 } \hfill \\
 {\lambda _1 ,\lambda _2 ,\lambda _3 } \hfill \\
\end{array} }} \right)}\cdot {\rm {\bf \upsilon }}+{\rm {\bf \upsilon
}}\cdot \nabla ^{\left( {{\begin{array}{*{20}c}
 {D_1 ,D_2 ,D_3 } \hfill \\
 {\lambda _1 ,\lambda _2 ,\lambda _3 } \hfill \\
\end{array} }} \right)}} \right).
\end{equation}
From $D_0 =D_1 =D_2 =D_3 =1$ and $\lambda _0 =\lambda _1 =\lambda _2
=\lambda _3 =1$, (\ref{eq65}) becomes the Cauchy strain tensor \cite{36}, and can be
applied to describe the power-law strain \cite{37}.

The Stokes-type strain tensor for the scaling-law flow, denoted by$\omega $,
is defined as
\begin{equation}
\label{eq66}
\omega =\frac{1}{2}\left( {\nabla ^{\left( {{\begin{array}{*{20}c}
 {D_1 ,D_2 ,D_3 } \hfill \\
 {\lambda _1 ,\lambda _2 ,\lambda _3 } \hfill \\
\end{array} }} \right)}\cdot {\rm {\bf \upsilon }}-{\rm {\bf \upsilon
}}\cdot \nabla ^{\left( {{\begin{array}{*{20}c}
 {D_1 ,D_2 ,D_3 } \hfill \\
 {\lambda _1 ,\lambda _2 ,\lambda _3 } \hfill \\
\end{array} }} \right)}} \right).
\end{equation}
The Stokes-type velocity gradient tensor for the scaling-law flow, denoted
by $\nabla ^{\left( {{\begin{array}{*{20}c}
 {D_1 ,D_2 ,D_3 } \hfill \\
 {\lambda _1 ,\lambda _2 ,\lambda _3 } \hfill \\
\end{array} }} \right)}\cdot {\rm {\bf \upsilon }}$, is presented as
follows:
\begin{equation}
\label{eq67}
\begin{array}{l}
 \nabla ^{\left( {{\begin{array}{*{20}c}
 {D_1 ,D_2 ,D_3 } \hfill \\
 {\lambda _1 ,\lambda _2 ,\lambda _3 } \hfill \\
\end{array} }} \right)}\cdot {\rm {\bf \upsilon }}=\omega +\Lambda \\
 =\frac{1}{2}\left( {\nabla ^{\left( {{\begin{array}{*{20}c}
 {D_1 ,D_2 ,D_3 } \hfill \\
 {\lambda _1 ,\lambda _2 ,\lambda _3 } \hfill \\
\end{array} }} \right)}\cdot {\rm {\bf \upsilon }}+{\rm {\bf \upsilon
}}\cdot \nabla ^{\left( {{\begin{array}{*{20}c}
 {D_1 ,D_2 ,D_3 } \hfill \\
 {\lambda _1 ,\lambda _2 ,\lambda _3 } \hfill \\
\end{array} }} \right)}} \right)\\
+\frac{1}{2}\left( {\nabla ^{\left(
{{\begin{array}{*{20}c}
 {D_1 ,D_2 ,D_3 } \hfill \\
 {\lambda _1 ,\lambda _2 ,\lambda _3 } \hfill \\
\end{array} }} \right)}\cdot {\rm {\bf \upsilon }}-{\rm {\bf \upsilon
}}\cdot \nabla ^{\left( {{\begin{array}{*{20}c}
 {D_1 ,D_2 ,D_3 } \hfill \\
 {\lambda _1 ,\lambda _2 ,\lambda _3 } \hfill \\
\end{array} }} \right)}} \right). \\
 \end{array}
\end{equation}
The stress tensor for the scaling-law flow, denoted by ${\rm {\bf U}}$, is
defined as
\begin{equation}
\label{eq68}
{\rm {\bf U}}=-p{\rm {\bf I}}+2\mu {\rm {\bf h}},
\end{equation}
where $\mu $ is the shear moduli of the viscosity, and ${\rm {\bf I}}$ is
the unit tensor.

Here, (\ref{eq66}) and (\ref{eq67}) are the generalized cases of the Stokes strain tensor
and Stokes velocity gradient tensor \cite{33}, when $D_0 =D_1 =D_2 =D_3 =1$ and
$\lambda _0 =\lambda _1 =\lambda _2 =\lambda _3 =1$.

The conservation of the momentums for the scaling-law flow is given as
follows:
\begin{equation}
\label{eq69}
\frac{D}{Dt}\mathop{\int\!\!\!\int\!\!\!\int}\limits_{\kern-5.5pt {\Omega
\left( t \right)}} {\rho {\rm {\bf \upsilon }}dV}
=\mathop{\int\!\!\!\int\!\!\!\int}\limits_{\kern-5.5pt {\Omega \left( t
\right)}} {{\rm {\bf W}}dV} +\mathop{{\int\!\!\!\!\!\int}\mkern-21mu
\bigcirc}\limits_{{\rm {\bf S}}\left( t \right)} {{\rm {\bf U}}\cdot {\rm
{\bf dS}}}
\end{equation}
where ${\rm {\bf W}}$ represents the specific body force.

Therefore, we have that
\begin{equation}
\label{eq70}
\left( {\lambda _0 D_0 t^{D_0 -1}} \right)\partial _t^{\left( 1 \right)}
\left( {\rho {\rm {\bf \upsilon }}} \right)+{\rm {\bf \upsilon }}\cdot
\nabla ^{\left( {{\begin{array}{*{20}c}
 {D_1 ,D_2 ,D_3 } \hfill \\
 {\lambda _1 ,\lambda _2 ,\lambda _3 } \hfill \\
\end{array} }} \right)}\left( {\rho {\rm {\bf \upsilon }}} \right)=\nabla
^{\left( {{\begin{array}{*{20}c}
 {D_1 ,D_2 ,D_3 } \hfill \\
 {\lambda _1 ,\lambda _2 ,\lambda _3 } \hfill \\
\end{array} }} \right)}\cdot {\rm {\bf U}}+{\rm {\bf W}}
\end{equation}
since
\begin{equation}
\label{eq71}
\begin{array}{l}
 \mathop{\int\!\!\!\int\!\!\!\int}\limits_{\kern-5.5pt {\Omega \left( t
\right)}} {\left( {\left( {\lambda _0 D_0 t^{D_0 -1}} \right)\partial
_t^{\left( 1 \right)} \left( {\rho {\rm {\bf \upsilon }}} \right)+{\rm {\bf
\upsilon }}\cdot \nabla ^{\left( {{\begin{array}{*{20}c}
 {D_1 ,D_2 ,D_3 } \hfill \\
 {\lambda _1 ,\lambda _2 ,\lambda _3 } \hfill \\
\end{array} }} \right)}\left( {\rho {\rm {\bf \upsilon }}} \right)}
\right)dV} \\
 =\mathop{\int\!\!\!\int\!\!\!\int}\limits_{\kern-5.5pt {\Omega \left( t
\right)}} {\left( {{\rm {\bf W}}+\nabla ^{\left( {{\begin{array}{*{20}c}
 {D_1 ,D_2 ,D_3 } \hfill \\
 {\lambda _1 ,\lambda _2 ,\lambda _3 } \hfill \\
\end{array} }} \right)}\cdot {\rm {\bf U}}} \right)dV} ,\\
 \end{array}
\end{equation}
where
\begin{equation}
\label{eq72}
\frac{D}{Dt}\mathop{\int\!\!\!\int\!\!\!\int}\limits_{\kern-5.5pt {\Omega
\left( t \right)}} {\rho {\rm {\bf \upsilon }}dV}
=\mathop{\int\!\!\!\int\!\!\!\int}\limits_{\kern-5.5pt {\Omega \left( t
\right)}} {\left[ {\left( {\lambda _0 D_0 t^{D_0 -1}} \right)\partial
_t^{\left( 1 \right)} \left( {\rho {\rm {\bf \upsilon }}} \right)+{\rm {\bf
\upsilon }}\cdot \nabla ^{\left( {{\begin{array}{*{20}c}
 {D_1 ,D_2 ,D_3 } \hfill \\
 {\lambda _1 ,\lambda _2 ,\lambda _3 } \hfill \\
\end{array} }} \right)}\left( {\rho {\rm {\bf \upsilon }}} \right)}
\right]dV}
\end{equation}
and
\begin{equation}
\label{eq73}
\mathop{{\int\!\!\!\!\!\int}\mkern-21mu \bigcirc}\limits_{{\rm {\bf
S}}\left( t \right)} {{\rm {\bf U}}\cdot {\rm {\bf dS}}}
=\mathop{\int\!\!\!\int\!\!\!\int}\limits_{\kern-5.5pt {\Omega \left( t
\right)}} {\nabla ^{\left( {{\begin{array}{*{20}c}
 {D_1 ,D_2 ,D_3 } \hfill \\
 {\lambda _1 ,\lambda _2 ,\lambda _3 } \hfill \\
\end{array} }} \right)}\cdot {\rm {\bf U}}dV} .
\end{equation}
From (\ref{eq68}) we have
\begin{equation}
\label{eq74}
\nabla ^{\left( {{\begin{array}{*{20}c}
 {D_1 ,D_2 ,D_3 } \hfill \\
 {\lambda _1 ,\lambda _2 ,\lambda _3 } \hfill \\
\end{array} }} \right)}\cdot {\rm {\bf U}}=-\nabla ^{\left(
{{\begin{array}{*{20}c}
 {D_1 ,D_2 ,D_3 } \hfill \\
 {\lambda _1 ,\lambda _2 ,\lambda _3 } \hfill \\
\end{array} }} \right)}p+\mu \nabla ^{\left( {{\begin{array}{*{20}c}
 {2D_1 ,2D_2 ,2D_3 } \hfill \\
 {\lambda _1 ,\lambda _2 ,\lambda _3 } \hfill \\
\end{array} }} \right)}{\rm {\bf \upsilon }}
\end{equation}
such that
\begin{equation}
\label{eq75}
\begin{array}{l}
\mathop{{\int\!\!\!\!\!\int}\mkern-21mu \bigcirc}\limits_{{\rm {\bf
S}}\left( t \right)} {{\rm {\bf U}}\cdot {\rm {\bf dS}}}
=\mathop{\int\!\!\!\int\!\!\!\int}\limits_{\kern-5.5pt {\Omega \left( t
\right)}} {\nabla ^{\left( {{\begin{array}{*{20}c}
 {D_1 ,D_2 ,D_3 } \hfill \\
 {\lambda _1 ,\lambda _2 ,\lambda _3 } \hfill \\
\end{array} }} \right)}\cdot {\rm {\bf U}}dV}\\
=-\mathop{\int\!\!\!\int\!\!\!\int}\limits_{\kern-5.5pt {\Omega \left( t
\right)}} {\nabla ^{\left( {{\begin{array}{*{20}c}
 {D_1 ,D_2 ,D_3 } \hfill \\
 {\lambda _1 ,\lambda _2 ,\lambda _3 } \hfill \\
\end{array} }} \right)}pdV}
+\mathop{\int\!\!\!\int\!\!\!\int}\limits_{\kern-5.5pt {\Omega \left( t
\right)}} {\mu \nabla ^{\left( {{\begin{array}{*{20}c}
 {2D_1 ,2D_2 ,2D_3 } \hfill \\
 {\lambda _1 ,\lambda _2 ,\lambda _3 } \hfill \\
\end{array} }} \right)}{\rm {\bf \upsilon }}dV} .
\end{array}
\end{equation}
It follows from (\ref{eq75}) that
\begin{equation}
\label{eq76}
\begin{array}{l}
 \frac{D}{Dt}\mathop{\int\!\!\!\int\!\!\!\int}\limits_{\kern-5.5pt {\Omega
\left( t \right)}} {(\rho {\rm {\bf \upsilon }})dV}
=\mathop{\int\!\!\!\int\!\!\!\int}\limits_{\kern-5.5pt {\Omega \left( t
\right)}} {{\rm {\bf W}}dV} +\mathop{{\int\!\!\!\!\!\int}\mkern-21mu
\bigcirc}\limits_{{\rm {\bf S}}\left( t \right)} {{\rm {\bf U}}\cdot {\rm
{\bf dS}}} \\
 =-\mathop{\int\!\!\!\int\!\!\!\int}\limits_{\kern-5.5pt {\Omega \left( t
\right)}} {\nabla ^{\left( {{\begin{array}{*{20}c}
 {D_1 ,D_2 ,D_3 } \hfill \\
 {\lambda _1 ,\lambda _2 ,\lambda _3 } \hfill \\
\end{array} }} \right)}pdV}
+\mathop{\int\!\!\!\int\!\!\!\int}\limits_{\kern-5.5pt {\Omega \left( t
\right)}} {\mu \nabla ^{\left( {{\begin{array}{*{20}c}
 {2D_1 ,2D_2 ,2D_3 } \hfill \\
 {\lambda _1 ,\lambda _2 ,\lambda _3 } \hfill \\
\end{array} }} \right)}{\rm {\bf \upsilon }}dV}
+\mathop{\int\!\!\!\int\!\!\!\int}\limits_{\kern-5.5pt {\Omega \left( t
\right)}} {{\rm {\bf W}}dV} . \\
 \end{array}
\end{equation}
From (\ref{eq63}) we obtain
\begin{equation}
\label{eq77}
\begin{array}{l}
 \frac{D}{Dt}\mathop{\int\!\!\!\int\!\!\!\int}\limits_{\kern-5.5pt {\Omega
\left( t \right)}} ({\rho {\rm {\bf \upsilon }}) dV} \\
 =\mathop{\int\!\!\!\int\!\!\!\int}\limits_{\kern-5.5pt {\Omega \left( t
\right)}} {\left[ {\left( {\lambda _0 D_0 t^{D_0 -1}} \right)\partial
_t^{\left( 1 \right)} (\rho \upsilon) +{\rm {\bf \upsilon }}\cdot \nabla ^{\left(
{{\begin{array}{*{20}c}
 {D_1 ,D_2 ,D_3 } \hfill \\
 {\lambda _1 ,\lambda _2 ,\lambda _3 } \hfill \\
\end{array} }} \right)}(\rho \upsilon) } \right]dV} \\
 =-\mathop{\int\!\!\!\int\!\!\!\int}\limits_{\kern-5.5pt {\Omega \left( t
\right)}} {\nabla ^{\left( {{\begin{array}{*{20}c}
 {D_1 ,D_2 ,D_3 } \hfill \\
 {\lambda _1 ,\lambda _2 ,\lambda _3 } \hfill \\
\end{array} }} \right)}pdV}
+\mathop{\int\!\!\!\int\!\!\!\int}\limits_{\kern-5.5pt {\Omega \left( t
\right)}} {\mu \nabla ^{\left( {{\begin{array}{*{20}c}
 {2D_1 ,2D_2 ,2D_3 } \hfill \\
 {\lambda _1 ,\lambda _2 ,\lambda _3 } \hfill \\
\end{array} }} \right)}{\rm {\bf \upsilon }}dV}
+\mathop{\int\!\!\!\int\!\!\!\int}\limits_{\kern-5.5pt {\Omega \left( t
\right)}} {{\rm {\bf W}}dV} \\
 =\mathop{\int\!\!\!\int\!\!\!\int}\limits_{\kern-5.5pt {\Omega \left( t
\right)}} {\left[ {-\nabla ^{\left( {{\begin{array}{*{20}c}
 {D_1 ,D_2 ,D_3 } \hfill \\
 {\lambda _1 ,\lambda _2 ,\lambda _3 } \hfill \\
\end{array} }} \right)}p+\mu \nabla ^{\left( {{\begin{array}{*{20}c}
 {2D_1 ,2D_2 ,2D_3 } \hfill \\
 {\lambda _1 ,\lambda _2 ,\lambda _3 } \hfill \\
\end{array} }} \right)}{\rm {\bf \upsilon }}+{\rm {\bf W}}} \right]dV} . \\
 \end{array}
\end{equation}
Therefore, from (\ref{eq63}) we present
\begin{equation}
\label{eq78}
\begin{array}{l}
 \mathop{\int\!\!\!\int\!\!\!\int}\limits_{\kern-5.5pt {\Omega \left( t
\right)}} {\left[ {\left( {\lambda _0 D_0 t^{D_0 -1}} \right)\partial
_t^{\left( 1 \right)} (\rho \upsilon) +{\rm {\bf \upsilon }}\cdot \nabla ^{\left(
{{\begin{array}{*{20}c}
 {D_1 ,D_2 ,D_3 } \hfill \\
 {\lambda _1 ,\lambda _2 ,\lambda _3 } \hfill \\
\end{array} }} \right)}(\rho \upsilon) } \right]dV} \\
 =\mathop{\int\!\!\!\int\!\!\!\int}\limits_{\kern-5.5pt {\Omega \left( t
\right)}} {\left[ {-\nabla ^{\left( {{\begin{array}{*{20}c}
 {D_1 ,D_2 ,D_3 } \hfill \\
 {\lambda _1 ,\lambda _2 ,\lambda _3 } \hfill \\
\end{array} }} \right)}p+\mu \nabla ^{\left( {{\begin{array}{*{20}c}
 {2D_1 ,2D_2 ,2D_3 } \hfill \\
 {\lambda _1 ,\lambda _2 ,\lambda _3 } \hfill \\
\end{array} }} \right)}{\rm {\bf \upsilon }}+{\rm {\bf W}}} \right]dV}, \\
\end{array}
\end{equation}
which leads to
\begin{equation}
\label{eq79}
\begin{array}{l}
\left( {\lambda _0 D_0 t^{D_0 -1}} \right)\partial _t^{\left( 1 \right)}
\left( {\rho {\rm {\bf \upsilon }}} \right)+{\rm {\bf \upsilon }}\cdot
\nabla ^{\left( {{\begin{array}{*{20}c}
 {D_1 ,D_2 ,D_3 } \hfill \\
 {\lambda _1 ,\lambda _2 ,\lambda _3 } \hfill \\
\end{array} }} \right)}\left( {\rho {\rm {\bf \upsilon }}} \right)=-\nabla
^{\left( {{\begin{array}{*{20}c}
 {D_1 ,D_2 ,D_3 } \hfill \\
 {\lambda _1 ,\lambda _2 ,\lambda _3 } \hfill \\
\end{array} }} \right)}p\\
+\mu \nabla ^{\left( {{\begin{array}{*{20}c}
 {2D_1 ,2D_2 ,2D_3 } \hfill \\
 {\lambda _1 ,\lambda _2 ,\lambda _3 } \hfill \\
\end{array} }} \right)}{\rm {\bf \upsilon }}+{\rm {\bf W}}.
\end{array}{l}
\end{equation}
In view of (\ref{eq79}), we obtain
\begin{equation}
\label{eq80}
\begin{array}{l}
\rho \left( {\left( {\lambda _0 D_0 t^{D_0 -1}} \right)\partial _t^{\left( 1
\right)} {\rm {\bf \upsilon }}+{\rm {\bf \upsilon }}\cdot \nabla ^{\left(
{{\begin{array}{*{20}c}
 {D_1 ,D_2 ,D_3 } \hfill \\
 {\lambda _1 ,\lambda _2 ,\lambda _3 } \hfill \\
\end{array} }} \right)}{\rm {\bf \upsilon }}} \right)\\
=-\nabla ^{\left(
{\alpha ,\beta ,\gamma } \right)}p+\mu \nabla ^{\left(
{{\begin{array}{*{20}c}
 {2D_1 ,2D_2 ,2D_3 } \hfill \\
 {\lambda _1 ,\lambda _2 ,\lambda _3 } \hfill \\
\end{array} }} \right)}{\rm {\bf \upsilon }}+{\rm {\bf W}},
\end{array}
\end{equation}
which yields that
\begin{equation}
\label{eq81}
\begin{array}{l}
\left( {\lambda _0 D_0 t^{D_0 -1}} \right)\partial _t^{\left( 1 \right)}
{\rm {\bf \upsilon }}+{\rm {\bf \upsilon }}\cdot \nabla ^{\left(
{{\begin{array}{*{20}c}
 {D_1 ,D_2 ,D_3 } \hfill \\
 {\lambda _1 ,\lambda _2 ,\lambda _3 } \hfill \\
\end{array} }} \right)}{\rm {\bf \upsilon }}\\
=-\frac{1}{\rho }\nabla ^{\left(
{\alpha ,\beta ,\gamma } \right)}p+\frac{\mu }{\rho }\nabla ^{\left(
{{\begin{array}{*{20}c}
 {2D_1 ,2D_2 ,2D_3 } \hfill \\
 {\lambda _1 ,\lambda _2 ,\lambda _3 } \hfill \\
\end{array} }} \right)}{\rm {\bf \upsilon }}+\frac{{\rm {\bf W}}}{\rho }.
\end{array}
\end{equation}
From (\ref{eq62}) we give
\begin{equation}
\label{eq82}
\nabla ^{\left( {{\begin{array}{*{20}c}
 {D_1 ,D_2 ,D_3 } \hfill \\
 {\lambda _1 ,\lambda _2 ,\lambda _3 } \hfill \\
\end{array} }} \right)}\cdot {\rm {\bf \upsilon }}=0.
\end{equation}
Thus, the Navier-Stokes-type equations of the scaling-law flows can be
written as follows:
\begin{equation}
\label{eq83}
\left\{ {\begin{array}{l}
 \left( {\lambda _0 D_0 t^{D_0 -1}} \right)\partial _t^{\left( 1 \right)}
\left( {\rho {\rm {\bf \upsilon }}} \right)+{\rm {\bf \upsilon }}\cdot
\nabla ^{\left( {{\begin{array}{*{20}c}
 {D_1 ,D_2 ,D_3 } \hfill \\
 {\lambda _1 ,\lambda _2 ,\lambda _3 } \hfill \\
\end{array} }} \right)}\left( {\rho {\rm {\bf \upsilon }}} \right)\\
=-\nabla
^{\left( {{\begin{array}{*{20}c}
 {D_1 ,D_2 ,D_3 } \hfill \\
 {\lambda _1 ,\lambda _2 ,\lambda _3 } \hfill \\
\end{array} }} \right)}p+\mu \nabla ^{\left( {{\begin{array}{*{20}c}
 {2D_1 ,2D_2 ,2D_3 } \hfill \\
 {\lambda _1 ,\lambda _2 ,\lambda _3 } \hfill \\
\end{array} }} \right)}{\rm {\bf \upsilon }}+{\rm {\bf W}}, \\
 \nabla ^{\left( {{\begin{array}{*{20}c}
 {D_1 ,D_2 ,D_3 } \hfill \\
 {\lambda _1 ,\lambda _2 ,\lambda _3 } \hfill \\
\end{array} }} \right)}\cdot {\rm {\bf \upsilon }}=0, \\
 \end{array}} \right.
\end{equation}
or
\begin{equation}
\label{eq84}
\left\{ {\begin{array}{l}
 \left( {\lambda _0 D_0 t^{D_0 -1}} \right)\partial _t^{\left( 1 \right)}
{\rm {\bf \upsilon }}+{\rm {\bf \upsilon }}\cdot \nabla ^{\left(
{{\begin{array}{*{20}c}
 {D_1 ,D_2 ,D_3 } \hfill \\
 {\lambda _1 ,\lambda _2 ,\lambda _3 } \hfill \\
\end{array} }} \right)}{\rm {\bf \upsilon }}\\
=-\frac{1}{\rho }\nabla ^{\left(
{\alpha ,\beta ,\gamma } \right)}p+\frac{\mu }{\rho }\nabla ^{\left(
{{\begin{array}{*{20}c}
 {2D_1 ,2D_2 ,2D_3 } \hfill \\
 {\lambda _1 ,\lambda _2 ,\lambda _3 } \hfill \\
\end{array} }} \right)}{\rm {\bf \upsilon }}+\frac{{\rm {\bf W}}}{\rho }, \\
 \nabla ^{\left( {{\begin{array}{*{20}c}
 {D_1 ,D_2 ,D_3 } \hfill \\
 {\lambda _1 ,\lambda _2 ,\lambda _3 } \hfill \\
\end{array} }} \right)}\cdot {\rm {\bf \upsilon }}=0. \\
 \end{array}} \right.
\end{equation}
On putting $D_0 =D_1 =D_2 =D_3 =1$ and $\lambda _0 =\lambda _1 =\lambda _2
=\lambda _3 =1$ in (\ref{eq83}) and (\ref{eq84}), we obtain the Navier-Stokes equations of
the scaling-law flow \cite{33,38}.

\section{Conclusion}

In the present study, the scaling-law vector calculus, which is connected
between the vector calculus and fractal geometry, was proposed due to the
calculus with respect to monotone functions. The Gauss, Ostrogradsky, Stokes
and Green tasks were extended based on the scaling-law vector calculus.
Making use of the material scaling-law derivative and transport theorem of
the scaling-law flows, the conservations of the mass and momentums for the
scaling-law flow were considered, and the Navier-Stokes-type equations of
the scaling-law flows were discussed in detail. The obtained formulas are
efficient and accurate for solving the challenge for the scaling-law flows.

\textbf{Acknowledgments}
This work is supported by the Yue-Qi Scholar of the China University of Mining and Technology (No. 102504180004).

\end{document}